\documentclass[wssquare]{ws-procs961x669} 
\begin{document}
\title{Effective black hole interior and the Raychadhuri equation}

\author{Keagan Blanchette}

\address{Department of Physics and
	Astronomy, York University~\\
	4700 Keele Street, Toronto, Ontario M3J 1P3 Canada\\
   E-mail: kblanch@yorku.ca}

\author{Saurya Das}

\address{Theoretical Physics Group and Quantum Alberta, Department of Physics and Astronomy, University of Lethbridge, 4401 University Drive, Lethhbridge, Alberta T1K 3M4, Canada\\
	E-mail: saurya.das@uleth.ca}

\author{Samantha Hergott}

\address{Department of Physics and
	Astronomy, York University~\\
	4700 Keele Street, Toronto, Ontario M3J 1P3 Canada\\
	E-mail: sherrgs@yorku.ca}

\author{Saeed Rastgoo}

\address{Department of Physics and
	Astronomy, York University~\\
	4700 Keele Street, Toronto, Ontario M3J 1P3 Canada\\
	E-mail: srastgoo@yorku.ca}

\begin{abstract}
We show that loop quantum gravity effects leads to the finiteness
of expansion and its rate of change in the effective regime in the
interior of the Schwarzschild black hole. As a consequence the singularity is resolved. We find this in line with previous results about curvature scalar and strong curvature singularities in Kantowski-Sachs model which is isometric to Schwarzschild interior.
\end{abstract}


\bodymatter

\section{Introduction}

Singularities are well-known predictions of General Relativity (GR).
They are recognized as regions that geodesics can reach in finite
proper time but cannot be extended beyond them. Such geodesics are
called incomplete. This notion can be formulated in terms of the existence
of conjugate points using the Raychaudhuri equation \citep{Raychaudhuri:1953yv}.
The celebrated Hawking-Penrose singularity theorems prove that under
normal assumptions, all spacetime solutions of GR will have incomplete
geodesics, and will therefore be singular \citep{Penrose:1964wq,Hawking:1969sw,Raychaudhuri:1953yv}.
These objects, however, are in fact predictions beyond the domain
of applicability of GR. So there is a consensus among the gravitational
physics community that they should be regularized in a full theory
of quantum gravity. Although there is no such theory available yet,
nevertheless there are a few candidates with rigorous mathematical
structure with which we can investigate the question of singularity
resolution. One such candidate is loop quantum gravity (LQG) \citep{Thiemann:2007pyv},
which is a connection-based canonical framework.

Within LQG, there have been numerous studies of both the interior
and the full spacetime of black holes in four and lower dimensions
\citep{Ashtekar:2005qt,Bohmer:2007wi,Corichi:2015xia,Ashtekar:2018cay,Bojowald:2016vlj,Chiou:2008nm,Corichi:2015vsa,Gambini:2020nsf,Gambini:2009vp,Rastgoo:2013isa,Corichi:2016nkp,Morales-Tecotl:2018ugi,Olmedo:2017lvt,Modesto:2006mx}. 
These attempts were originally inspired by loop quantum cosmology
(LQC), more precisely a certain quantization of the isotropic Friedmann-Lemaitre-Robertson-Walker
(FLRW) model \citep{Ashtekar:2006rx,Ashtekar:2006uz} which uses a
certain type of quantization of the phase space variables called polymer
quantization \citep{Ashtekar:2002sn,Corichi:2007tf,Morales-Tecotl:2016ijb,Tecotl:2015cya,Flores-Gonzalez:2013zuk}.
This quantiztion introduces a so called polymer parameter that sets
the scale at which the quantum effects become important. There are
various schemes of such quantization based on the form of the polymer
parameter.

In this paper, we examine the issue of singularity resolution via
the LQG-modified Raychaudhuri equation for the interior of the Schwarzschild
black hole. By choosing adapted holonomies and fluxes, which are the
conjugate variables in LQG, and using polymer quantization, we compute
the corresponding expansion of geodesics and derive the effective
Raychaudhuri equation. This way we show effective terms introduce
a repulsive effect which prevents the formation of conjugate points.
This implies that the classical singularity theorems are rendered
invalid and the singularity is resolved, at least for the spacetime
under consideration. The boundedness of expansion scalar and the resolution of strong curvature singularities have been studied and shown in the context of Kantowski-Sachs model in \citep{Joe:2014tca,Saini:2016vgo}.

This paper is organized as follows. In Sec. \ref{sec:Schw-int}, we
review the classical interior of the Schwarzschild black hole. In
Sec. \ref{sec:Ray}, we briefly discuss the classical Raychaudhuri
equation and its importance. Then, in Sec. \ref{subsec:Class-Ray}
we present the behavior of the Raychaudhuri equation in the classical
regime. In Sec. \ref{sec:Effective-Dynamics}, the effective Raychaudhuri
equation for three different schemes of polymer quantization are derived
and are compared with the classical behavior. Finally, in Sec. \ref{Sec:Disc}
we briefly discuss our results and conclude.

\section{Interior of the Schwarzschild black hole\label{sec:Schw-int}}

It is well known that the metric of the interior of the Schwarzschild
black hole can be obtained by switching the the Schwarzschild coordinates
$t$ and $r$, due to the fact that spacelike and timelike curves
switch their causal nature upon crossing the horizon. This yields
the metric of the interior as 
\begin{equation}
ds^{2}=-\left(\frac{2GM}{t}-1\right)^{-1}dt^{2}+\left(\frac{2GM}{t}-1\right)dr^{2}+t^{2}\left(d\theta^{2}+\sin^{2}\theta d\phi^{2}\right).\label{eq:sch-inter}
\end{equation}
This metric is in fact a special case of a Kantowski-Sachs cosmological
spacetime \citep{Collins:1977fg} 
\begin{equation}
ds_{KS}^{2}=-N(T)^{2}dT^{2}+g_{xx}(T)dx^{2}+g_{\theta\theta}(T)d\theta^{2}+g_{\phi\phi}(T)d\phi^{2}\label{eq:K-S-gener}
\end{equation}

One can obtain the Hamiltonian of the interior system in connection
variables, one first considers the full Hamiltonian of gravity written
in terms of (the curvature) of the $su(2)$ Ashtekar-Barbero connection
$A_{a}^{i}$ and its conjugate momentum, the densitized triad $\tilde{E}_{a}^{i}$.
Using the Kantowski-Sachs symmetry, these variables can be written
as \citep{Ashtekar:2005qt} 
\begin{align}
A_{a}^{i}\tau_{i}dx^{a}= & \frac{c}{L_{0}}\tau_{3}dx+b\tau_{2}d\theta-b\tau_{1}\sin\theta d\phi+\tau_{3}\cos\theta d\phi,\label{eq:A-AB}\\
\tilde{E}_{i}^{a}\tau_{i}\partial_{a}= & p_{c}\tau_{3}\sin\theta\partial_{x}+\frac{p_{b}}{L_{0}}\tau_{2}\sin\theta\partial_{\theta}-\frac{p_{b}}{L_{0}}\tau_{1}\partial_{\phi},\label{eq:E-AB}
\end{align}
where $b$, $c$, $p_{b}$, and $p_{c}$ are functions that only depend
on time and $\tau_{i}=-i\sigma_{i}/2$ are a $su(2)$ basis satisfying
$\left[\tau_{i},\tau_{j}\right]=\epsilon_{ij}{}^{k}\tau_{k}$, with
$\sigma_{i}$ being the Pauli matrices. Here $L_{0}$ is a fiducial
length of a fiducial volume, chosen to restrict the integration limits
of the symplectic form so that the integral does not diverge. Substituting
these into the full Hamiltonian of gravity written in Ashtekar connection
variables, one obtains the symmetry reduced Hamiltonian constraint
adapted to this model as \citep{Ashtekar:2005qt,Chiou:2008nm,Corichi:2015xia,Bohmer:2007wi,Morales-Tecotl:2018ugi}
\begin{equation}
H=-\frac{N}{2G\gamma^{2}}\left[\left(b^{2}+\gamma^{2}\right)\frac{p_{b}}{\sqrt{p_{c}}}+2bc\sqrt{p_{c}}\right].\label{eq:H-const-class}
\end{equation}
while the diffeomorphism constraint vanishes identically due to the
homogenous nature of the model. Here, $\gamma$ is the Barbero-Immirzi
parameter \citep{Thiemann:2007pyv}, and $p_{c}\geq0$. The corresponding
Poisson brackets of the model become 
\begin{equation}
\{c,p_{c}\}=2G\gamma,\quad\quad\{b,p_{b}\}=G\gamma.\label{eq:classic-PBs-bc}
\end{equation}
The general form of the Kantowski-Sachs metric written in terms of
the above variables becomes 
\begin{equation}
ds^{2}=-N(T)^{2}dT^{2}+\frac{p_{b}^{2}(T)}{L_{0}^{2}\,p_{c}(T)}dx^{2}+p_{c}(T)(d\theta^{2}+\sin^{2}\theta d\phi^{2}).
\end{equation}
Comparing this with the standard Schwarzschild interior metric one
obtains 
\begin{align}
p_{b}= & 0, & p_{c}= & 4G^{2}M^{2}, &  & \textrm{on the horizon\,}t=2GM,\label{eq:t-horiz}\\
p_{b}\to & 0, & p_{c}\to & 0, &  & \textrm{at the singularity\,}t\to0.\label{eq:t-singular}
\end{align}

\section{The Raychaudhuri equation\label{sec:Ray}}

The celebrated Raychaudhuri equation \citep{Raychaudhuri:1953yv}
\begin{equation}
\frac{d\theta}{d\tau}=-\frac{1}{3}\theta^{2}-\sigma_{ab}\sigma^{ab}+\omega_{ab}\omega^{ab}-R_{ab}U^{a}U^{b}
\end{equation}
describes the behavior of geodesics in spacetime purely geometrically
and independent of the theory of gravity under consideration. Here,
$\theta$ is the expansion term describing how geodesics focus or
defocus; $\sigma_{ab}\sigma^{ab}$ is the shear which describes how,
e.g., a circular configuration of geodesics changes shape into, say,
an ellipse; $\omega_{ab}\omega^{ab}$ is the vorticity term; $R_{ab}$
is the Ricci tensor; and $U^{a}$ is the tangent vector to the geodesics.
Note that, due the sign of the expansion, shear, and the Ricci term,
they all contribute to focusing, while the vorticity terms leads to
defocusing.

In our case, since we consider the model in vacuum, $R_{ab}=0$. Also,
in general in Kantowski-Scahs models, the vorticity term is only nonvanishing
if one considers metric perturbations \citep{Collins:1977fg}. Hence,
$\omega_{ab}\omega^{ab}=0$ in our model, too. This reduces the Raychaudhuri
equation for our analysis to 
\begin{equation}
\frac{d\theta}{d\tau}=-\frac{1}{3}\theta^{2}-\sigma_{ab}\sigma^{ab}.\label{eq:Ray-1}
\end{equation}
To obtain the right hand side of the equation above, we need to consider
a congruence of geodesics and derive their expansion and shear. By
choosing such a congruence with 4-velocities $U^{a}=\left(\frac{1}{N},0,0,0\right)$
we obtain 
\begin{align}
\theta= & \frac{\dot{p}_{b}}{Np_{b}}+\frac{\dot{p}_{c}}{2Np_{c}},\label{eq:expansion}\\
\sigma^{2}= & \frac{2}{3}\left(-\frac{\dot{p}_{b}}{Np_{b}}+\frac{\dot{p}_{c}}{Np_{c}}\right)^{2}.\label{eq:shear}
\end{align}

\section{Classical vs effective Raychaudhuri equation}

\subsection{Classical Raychaudhuri equation\label{subsec:Class-Ray}}

\begin{figure}
\begin{centering}
\includegraphics[scale=0.4]{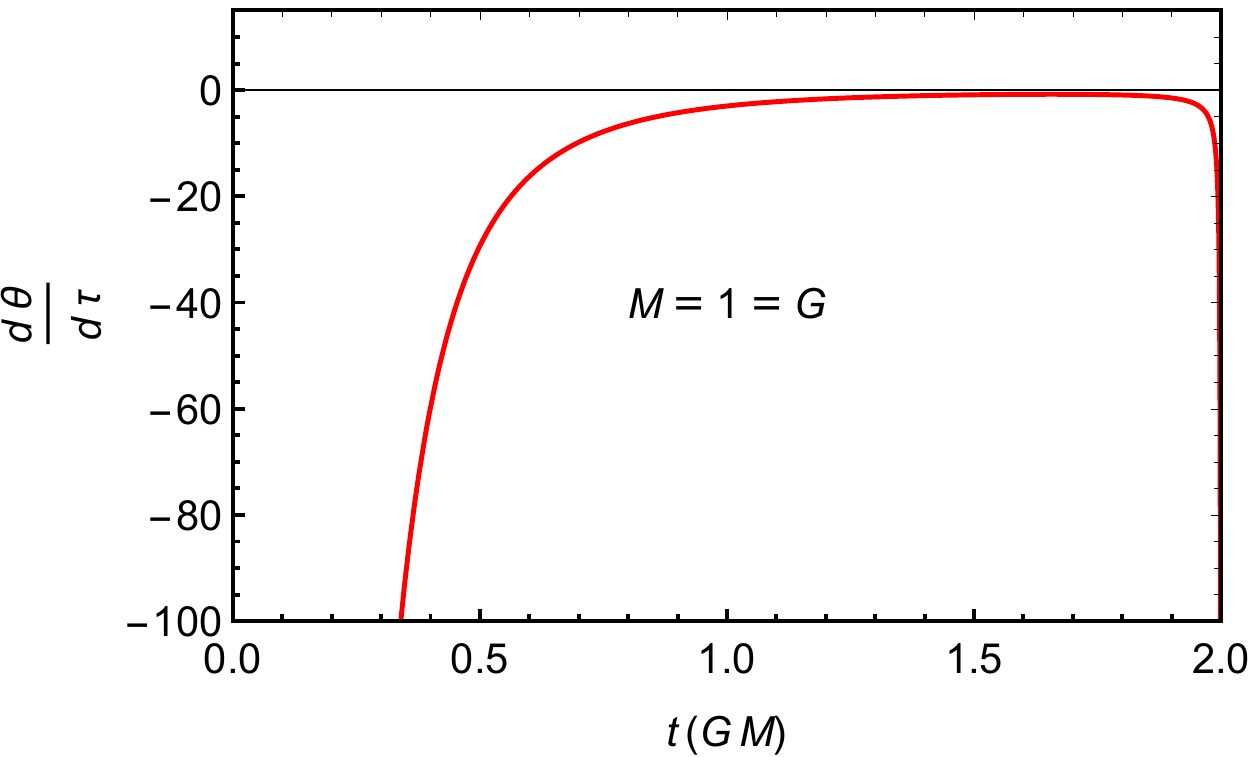}~~~~~~~~~~~\includegraphics[scale=0.4]{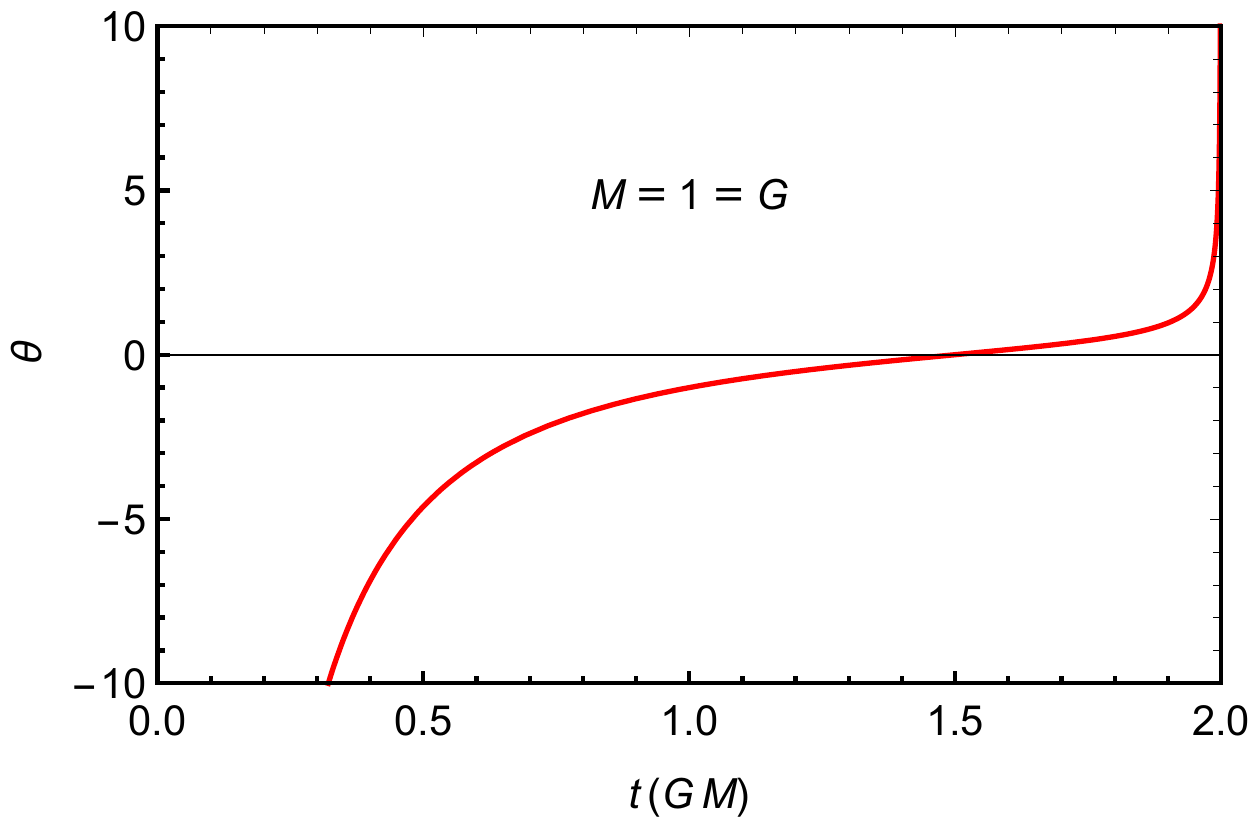} 
\par\end{centering}
\caption{Left: $\frac{d\theta}{d\tau}$ as a function of the Schwarzschild
time $t$. Right: negative branch of $\theta$ as a function of $t$.
Both $\theta$ and $\frac{d\theta}{d\tau}$ diverge as we approach
$t\to0$. Note that the divergence at the horizon is due to the choice
of Schwarzschild coordinate system. \label{fig:RE-cls-t}}
\end{figure}

Having obtained the adapted form of the Raychaudhuri equation for
our model, we set to find it explicitly. Looking at \eqref{eq:Ray-1}-\eqref{eq:shear}
we see that we need the solutions to the equations of motion to be
able to compute them. In order to facilitate such a derivation, we
choose a gauge where the lapse function is 
\begin{equation}
N\left(T\right)=\frac{\gamma\sqrt{p_{c}\left(T\right)}}{b\left(T\right)},\label{eq:lapsNT}
\end{equation}
for which the Hamiltonian constraint becomes 
\begin{equation}
H=-\frac{1}{2G\gamma}\left[\left(b^{2}+\gamma^{2}\right)\frac{p_{b}}{b}+2cp_{c}\right].\label{eq:H-const-cls-gauged}
\end{equation}
The advantage of this lapse function is that the equations of motion
of $c,\,p_{c}$ decouple from those of $b,\,p_{b}$. These equations
of motion should be solved together with enforcing the vanishing of
the Hamiltonian constraint \eqref{eq:H-const-cls-gauged} on-shell
(i.e., on the constraint surface). Replacing these solutions into
\eqref{eq:Ray-1} one obtains 
\begin{align}
\theta= & \pm\frac{1}{2\sqrt{p_{c}}}\left(\frac{3b}{\gamma}-\frac{\gamma}{b}\right)=\pm\frac{-2t+3GM}{t^{\frac{3}{2}}\sqrt{(2GM-t)}},\label{eq:theta-t-class}\\
\frac{d\theta}{d\tau}= & -\frac{1}{2p_{c}}\left(1+\frac{9b^{2}}{2\gamma^{2}}+\frac{\gamma^{2}}{2b^{2}}\right)=\frac{-2t^{2}+8GMt-9G^{2}M^{2}}{\left(2GM-t\right)t^{3}}.\label{eq:RE-RHS-t-class}
\end{align}
As expected, the right hand side of $\frac{d\theta}{d\tau}$ is negative
(since $p_{c}>0$) and both $\theta$ and $\frac{d\theta}{d\tau}$
diverge at the singularity in the classical regime. This can be seen
from Fig. \ref{fig:RE-cls-t} which reaffirm the existence of a classical
singularity at the center of the black hole.

\subsection{Effective dynamics and Raychaudhuri equation\label{sec:Effective-Dynamics}}

The effective behavior of the interior of the Schwarzschild black
hole can be deduced from its effective Hamiltonian (constraint). There
are various equivalent ways to obtain such an effective Hamiltonian
from the classical one \citep{Ashtekar:2005qt,Chiou:2008nm,Corichi:2015xia,Bohmer:2007wi,Morales-Tecotl:2018ugi}.
It turns out that the easiest way is by replacing 
\begin{align}
b\to & \frac{\sin\left(\mu_{b}b\right)}{\mu_{b}},\label{eq:b-to-sinb}\\
c\to & \frac{\sin\left(\mu_{c}c\right)}{\mu_{c}},\label{eq:c-to-sinc}
\end{align}
in the classical Hamiltonian.

The free parameters $\mu_{b},\,\mu_{c}$ are the minimum scales associated
with the radial and angular directions \citep{Ashtekar:2005qt,Corichi:2015xia,Chiou:2008nm,Chiou:2008eg}.
If these $\mu$ parameters are taken to be constant, the corresponding
approach is called the $\mu_{0}$ scheme. If, however, these parameters
depend on the conjugate momenta, the approach is called improved dynamics
which itself is divided into various subcategories. In case of the
Schwarzschild interior and due to lack of matter content, it is not
clear which scheme yields the correct semiclassical limit. Hence,
for completeness, in this paper we will study the effective theory
in the constant $\mu$ scheme, which here we call the $\mathring{\mu}$
scheme, as well as in two of the most common improved schemes, which
we denote by $\bar{\mu}$ and $\bar{\mu}^{\prime}$ schemes.

Replacing \eqref{eq:b-to-sinb} and \eqref{eq:c-to-sinc} into the
classical Hamiltonian \eqref{eq:H-const-class}, one obtains an effective
Hamiltonian constraint, 
\begin{equation}
H_{\textrm{eff}}^{(N)}=-\frac{N}{2G\gamma^{2}}\left[\left(\frac{\sin^{2}\left(\mu_{b}b\right)}{\mu_{b}^{2}}+\gamma^{2}\right)\frac{p_{b}}{\sqrt{p_{c}}}+2\frac{\sin\left(\mu_{b}b\right)}{\mu_{b}}\frac{\sin\left(\mu_{c}c\right)}{\mu_{c}}\sqrt{p_{c}}\right].\label{eq:H-eff-gen}
\end{equation}
In order to be able to compare the effective results with the classical
case, we need to use the same lapse as we did in the classical part.
Under \eqref{eq:b-to-sinb}, this lapse \eqref{eq:lapsNT} becomes
\begin{equation}
N=\frac{\gamma\mu_{b}\sqrt{p_{c}}}{\sin\left(\mu_{b}b\right)}.\label{eq:laps-eff}
\end{equation}
Using this in \eqref{eq:H-eff-gen} we obtain 
\begin{equation}
H_{\textrm{eff}}=-\frac{1}{2\gamma G}\left[p_{b}\left[\frac{\sin\left(\mu_{b}b\right)}{\mu_{b}}+\gamma^{2}\frac{\mu_{b}}{\sin\left(\mu_{b}b\right)}\right]+2p_{c}\frac{\sin\left(\mu_{c}c\right)}{\mu_{c}}\right].\label{eq:H-eff-spec}
\end{equation}
Note that both \eqref{eq:H-eff-gen} and \eqref{eq:H-eff-spec} reduce
to their classical counterparts \eqref{eq:H-const-class} and \eqref{eq:H-const-cls-gauged}
respectively, as is expected.

To obtain the effective Raychadhuri equation, we consider three cases:
\begin{enumerate}
\item \label{enu:mu-0}$\mathring{\mu}$ scheme where $\mu_{b}=\mathring{\mu}_{b}$
and $\mu_{c}=\mathring{\mu}_{c}$ are constants,
\item \label{enu:mu-bar}$\bar{\mu}$ scheme where we set
\begin{align}
\mu_{b}=\bar{\mu}_{b}= & \sqrt{\frac{\Delta}{p_{b}}},\label{eq:mu-bar-b}\\
\mu_{c}=\bar{\mu}_{c}= & \sqrt{\frac{\Delta}{p_{c}}},\label{eq:mu-bar-c}
\end{align}
\item \label{enu:mu-bar-prime}$\bar{\mu}^{\prime}$ scheme where we choose
\begin{align}
\mu_{b}=\bar{\mu}_{b}^{\prime}= & \sqrt{\frac{\Delta}{p_{c}}},\label{eq:mu-bar-b-prime}\\
\mu_{c}=\bar{\mu}_{c}^{\prime}= & \frac{\sqrt{p_{c}\Delta}}{p_{b}}.\label{eq:mu-bar-c-prime}
\end{align}
After replacing these (separately for each case) into the effective
Hamiltonian constraint \eqref{eq:H-eff-spec} and finding their corresponding
equations of motion \citep{Blanchette:2020kkk}, one replaces the
solutions in the Raychadhuri equation \eqref{eq:Ray-1} to obtain
the form of $\frac{d\theta}{d\tau}$. It turns out that for all the
three cases above we obtain
\begin{align}
\frac{d\theta}{d\tau}= & \frac{1}{\gamma^{2}p_{c}}\frac{\sin^{2}\left(\mu_{b}b\right)}{\mu_{b}^{2}}\left[\cos\left(\mu_{b}b\right)\cos\left(\mu_{c}c\right)-\frac{\cos^{2}\left(\mu_{b}b\right)}{4}-3\cos^{2}\left(\mu_{c}c\right)\right]\nonumber \\
 & +\frac{\cos\left(\mu_{b}b\right)}{p_{c}}\left[\frac{\cos\left(\mu_{b}b\right)}{2}-\cos\left(\mu_{c}c\right)-\frac{\gamma^{2}}{4}\cos\left(\mu_{b}b\right)\frac{\mu_{b}^{2}}{\sin^{2}\left(\mu_{b}b\right)}\right],\label{eq:dtheta-dtau-mu-all}
\end{align}
where it is understood that $\mu$'s should be substituted for from
cases \ref{enu:mu-0}--\ref{enu:mu-bar-prime} suitably for each
case.
\end{enumerate}

\subsubsection{$\mathring{\mu}$ scheme\label{subsec:mu0-scheme}}

\begin{figure}
\begin{centering}
\includegraphics[scale=0.5]{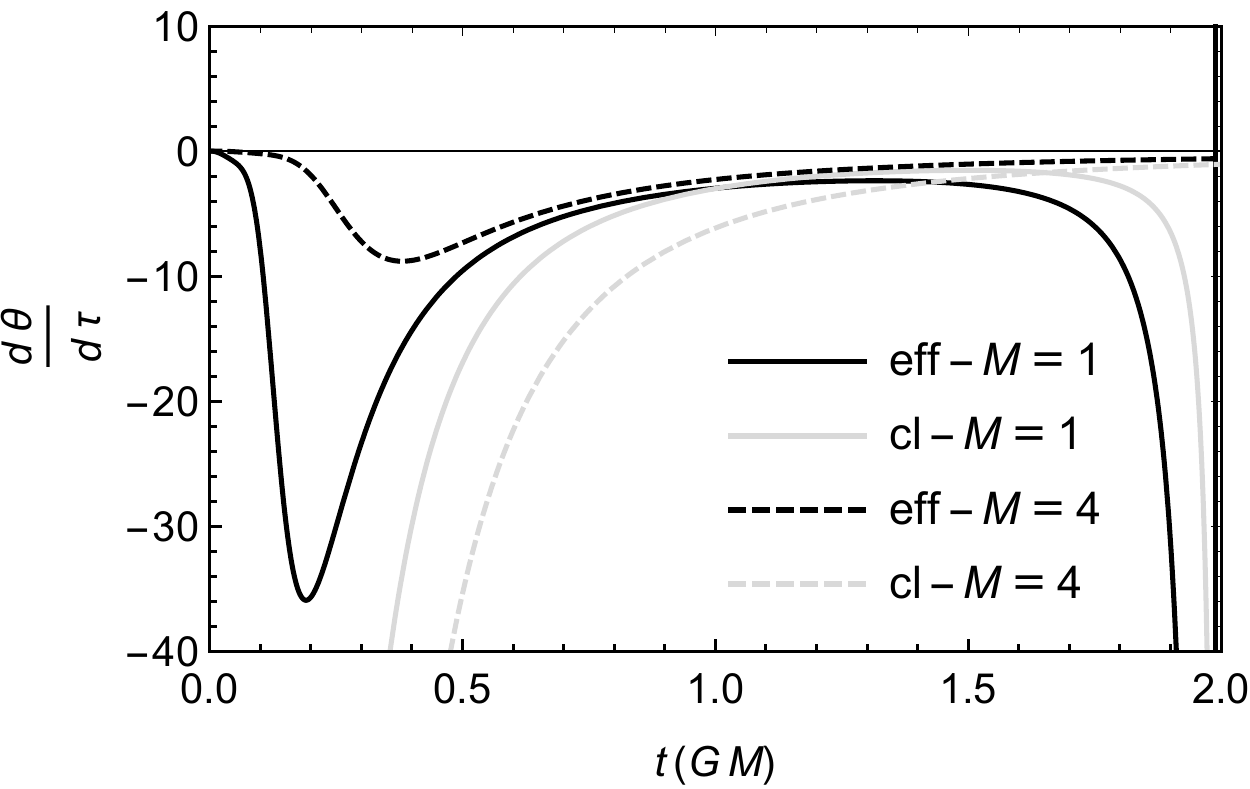} 
\par\end{centering}
\caption{Plot of $\frac{d\theta}{d\tau}$ as a function of the Schwarzschild
time $t$, for two different masses in classical vs effective regimes.
The figure is plotted using $\gamma=0.5,\,G=1,\,L_{0}=1$, and $\mathring{\mu}_{b}=0.08=\mathring{\mu}_{c}$.
\label{fig:RE-eff-mu0}}
\end{figure}

Let us first consider this case perturbatively in an analytic manner.
Replacing $\mu_{b}=\mathring{\mu}_{b}$ and $\mu_{c}=\mathring{\mu}_{c}$
as constants in \eqref{eq:dtheta-dtau-mu-all} and then expanding
for small values of $\mu$'s up to the second order we obtain
\begin{equation}
\frac{d\theta}{d\tau}\approx-\frac{1}{2p_{c}}\left(1+\frac{9b^{2}}{2\gamma^{2}}+\frac{\gamma^{2}}{2b^{2}}\right)+\mathring{\mu}_{b}^{2}\frac{1}{2p_{c}}\left(\frac{b^{4}}{\gamma^{2}}+\frac{\gamma^{2}}{3}\right)+\mathring{\mu}_{c}^{2}\frac{c^{2}}{2p_{c}}\left(1+\frac{5b^{2}}{\gamma^{2}}\right).\label{eq:dtheta-dtau-mu0-pert}
\end{equation}
It is seen that the first term on the right-hand side above is the
classical expression \eqref{eq:RE-RHS-t-class} which is always negative
and leads to the divergence of classical expansion rate at the singularity,
i.e., infinite focusing. However, Eq. \eqref{eq:dtheta-dtau-mu0-pert}
now involves two additional effective terms proportional to $\mathring{\mu}_{b}^{2}$
and $\mathring{\mu}_{c}^{2}$, both of which are positive. Furthermore,
from the solutions to equations of motion \citep{Blanchette:2020kkk},
one can infer that these two terms take over close to where the classical
singularity used to be and stop $\theta$ and $\frac{d\theta}{d\tau}$
from diverging. This can, in fact, be confirmed by looking at the
full nonperturbative behavior of $\frac{d\theta}{d\tau}$ plotted
in Fig. \ref{fig:RE-eff-mu0}. There, it is seen that the quantum
gravity effects counter the attractive nature of classical terms and
turn the curve around such that $\frac{d\theta}{d\tau}$ goes to zero
for $t\to0$.

\subsubsection{$\bar{\mu}$ scheme\label{subsec:mubar-1-scheme}}

Similar to the $\mathring{\mu}$ scheme we start by analyzing the
perturbative expansion of $\frac{d\theta}{d\tau}$ for this case by
replacing \eqref{eq:mu-bar-b} and \eqref{eq:mu-bar-c} in \eqref{eq:dtheta-dtau-mu-all}
and expanding it for small $\bar{\mu}$'s up to the lowest correction
terms which in this case is $\Delta$ (which can be considered as
the second order in $\bar{\mu}$ scales). This way we get 
\begin{equation}
\frac{d\theta}{d\tau}\approx-\frac{1}{2p_{c}}\left(1+\frac{9b^{2}}{2\gamma^{2}}+\frac{\gamma^{2}}{2b^{2}}\right)+\frac{\Delta}{p_{c}}\left[\frac{1}{6p_{b}}\left(\frac{3b^{4}}{\gamma^{2}}+\gamma^{2}\right)+\frac{c^{2}}{2p_{c}}\left(1+\frac{5b^{2}}{\gamma^{2}}\right)\right].
\end{equation}
Once again, the first term on the right-hand side is the classical
expression of the Raychaudhuri equation \eqref{eq:RE-RHS-t-class},
which contributes to infinite focusing at the singularity, but all
the correction terms are positive and take over close to the position
of the classical singularity. This stops $\frac{d\theta}{d\tau}$
from diverging similar to the $\mathring{\mu}$ scheme.

\begin{figure}
\begin{centering}
\includegraphics[scale=0.4]{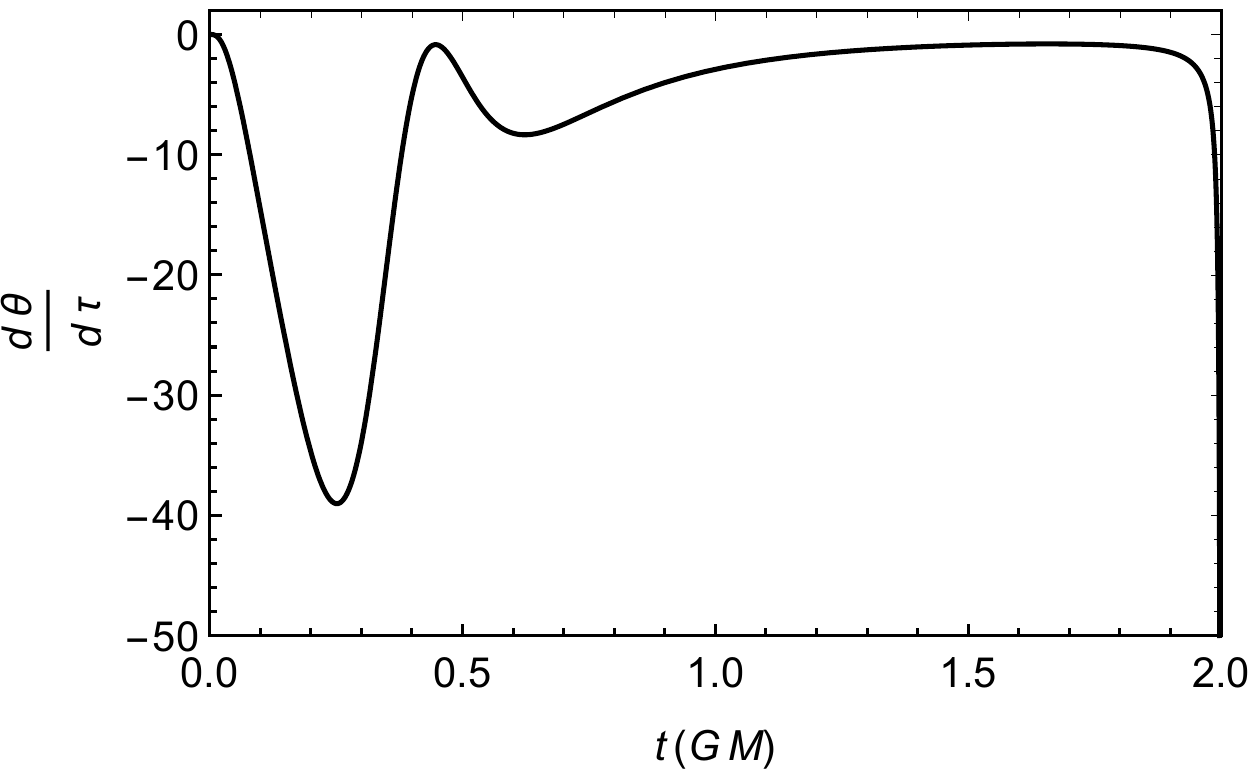}~~~~~~~~~~~~~~~~~\includegraphics[scale=0.4]{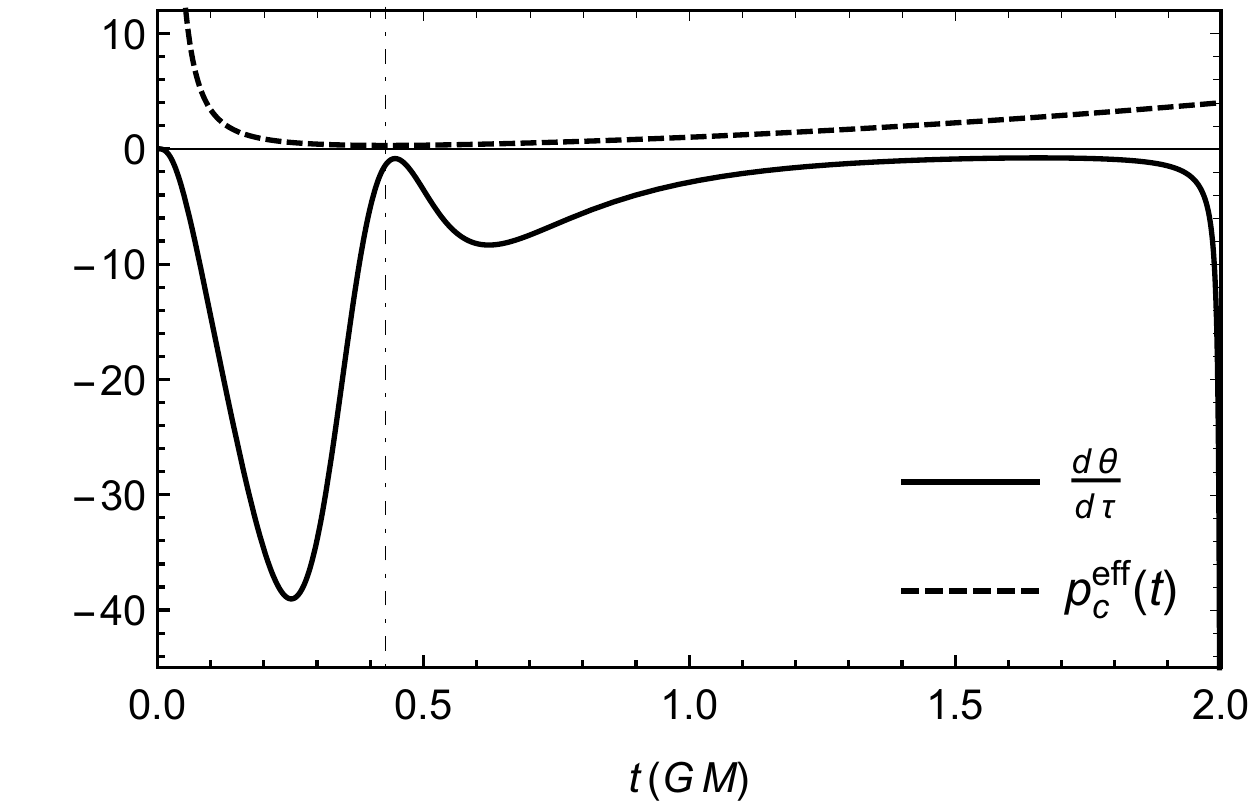} 
\par\end{centering}
\caption{Left: Raychaudhuri equation in the $\bar{\mu}$ scheme. Right: Raychaudhuri
equation vs $p_{c}$. The vertical dot-dashed line at $t\approx0.43GM$
is the position of the bounce of $p_{c}$ where its minimum $p_{c}^{\textrm{min}}=0.29$
happens in this case. The figure is plotted using $\gamma=0.5,\,M=1,\,G=1,\,L_{0}=1$,
and $\Delta=0.1$. \label{fig:RE-bar}}
\end{figure}

The full nonperturbative form of the modified Raychaudhuri equation
in terms of $t$ can also be plotted plotted by substituting the numerical
solutions of the equations of motion for the $\bar{\mu}$ in \eqref{eq:dtheta-dtau-mu-all}.
The result is plotted in Fig. \ref{fig:RE-bar}. We see that, approaching
from the horizon to where the classical singularity used to be, an
initial bump or bounce in encountered, followed by a more pronounced
bounce closer to where the singularity used to be. Once again, the
quantum corrections become dominant close to the singularity and turn
back the $\frac{d\theta}{d\tau}$ such that at $t\to0$ no focusing
happens at all. Furthermore, from the right plot in Fig. \ref{fig:RE-bar},
we see that the first bounce in the Raychaudhuri equation happens
much earlier than the bounce in $p_{c}$ for this batch of geodesics.


\subsubsection{$\bar{\mu}^{\prime}$ scheme\label{subsec:mubar-2-scheme}}

The perturbative analytical form of $\frac{d\theta}{d\tau}$ for this
case, up to the first correction term turns out to be 
\begin{equation}
\frac{d\theta}{d\tau}\approx-\frac{1}{2p_{c}}\left(1+\frac{9b^{2}}{2\gamma^{2}}+\frac{\gamma^{2}}{2b^{2}}\right)+\frac{\Delta}{6\gamma^{2}}\left[\frac{1}{p_{c}^{2}}\left(3b^{4}+\gamma^{4}\right)+\frac{3c^{2}}{p_{b}^{2}}\left(5b^{2}+\gamma^{2}\right)\right].
\end{equation}
Although this perturbative form of the Raychaudhuri equation is a
bit different from previous cases, nevertheless it exhibits the property
that the quantum corrections are all positive and take over close
to where the classical singularity used to be, and hence once again
contribute to defocusing of the geodesics. This case is, however,
rather different from the previous two cases since the behavior of
some of the canonical variables as a function of the Schwarzschild
time $t$ deviates significantly from those cases. In particular,
both $b$ and $p_{c}$ show a kind of damped oscillatory behavior
close to the classical singularity \citep{Blanchette:2020kkk}, which
contributes to a more volatile behavior of the Raychaudhuri equation.

\begin{figure}
\begin{centering}
\includegraphics[scale=0.4]{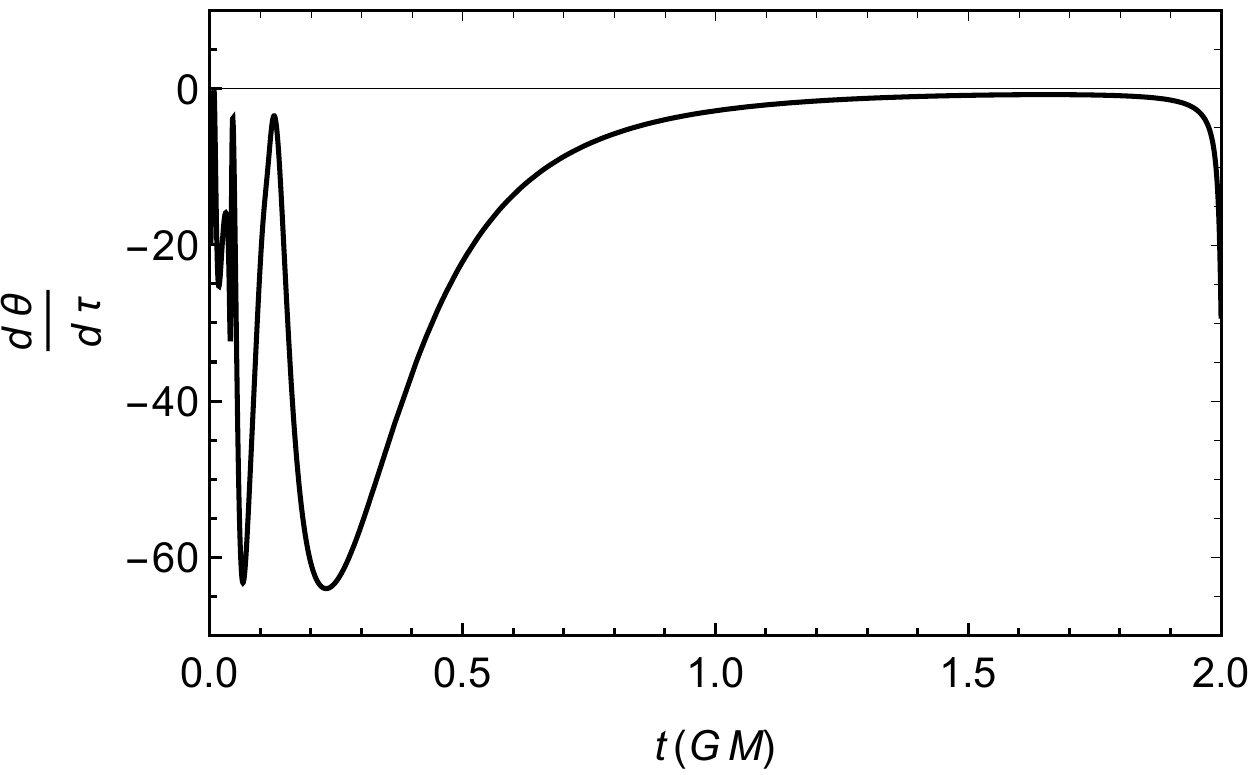}~~~~~~~~\includegraphics[scale=0.4]{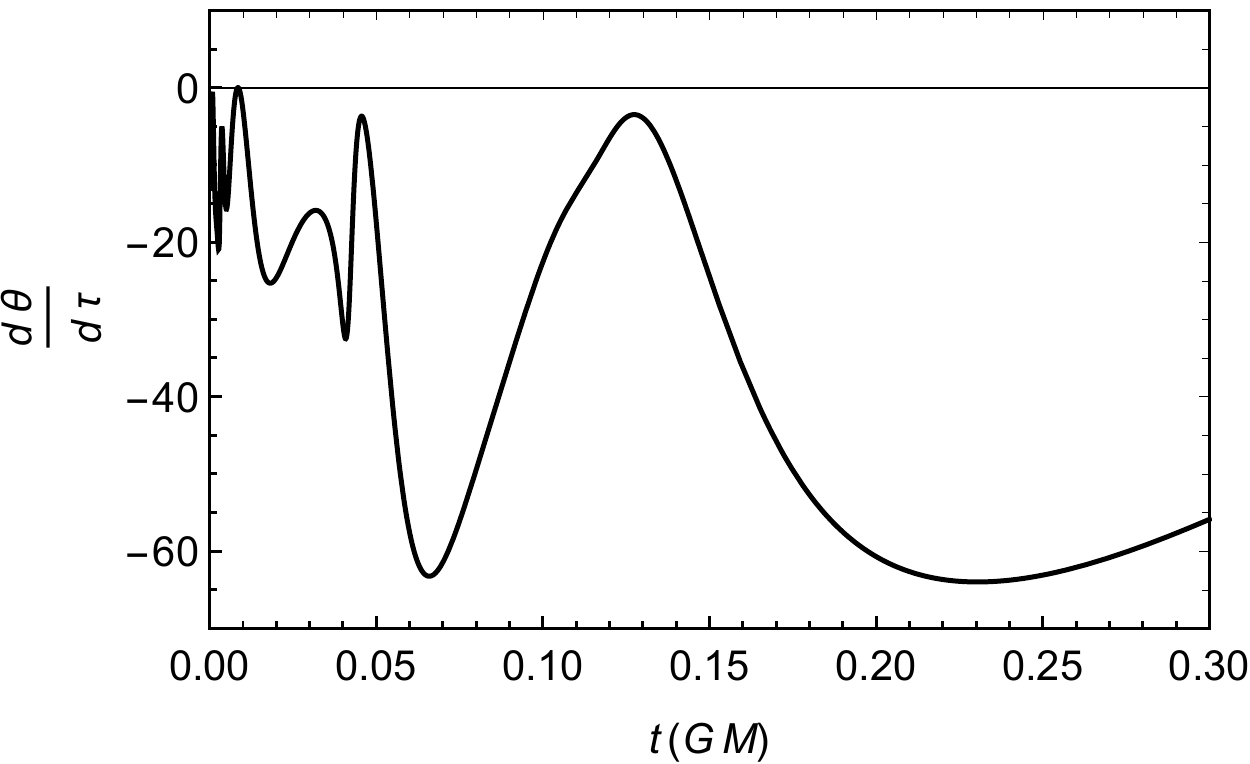} 
\par\end{centering}
\begin{centering}
\vspace{10pt}
 \includegraphics[scale=0.4]{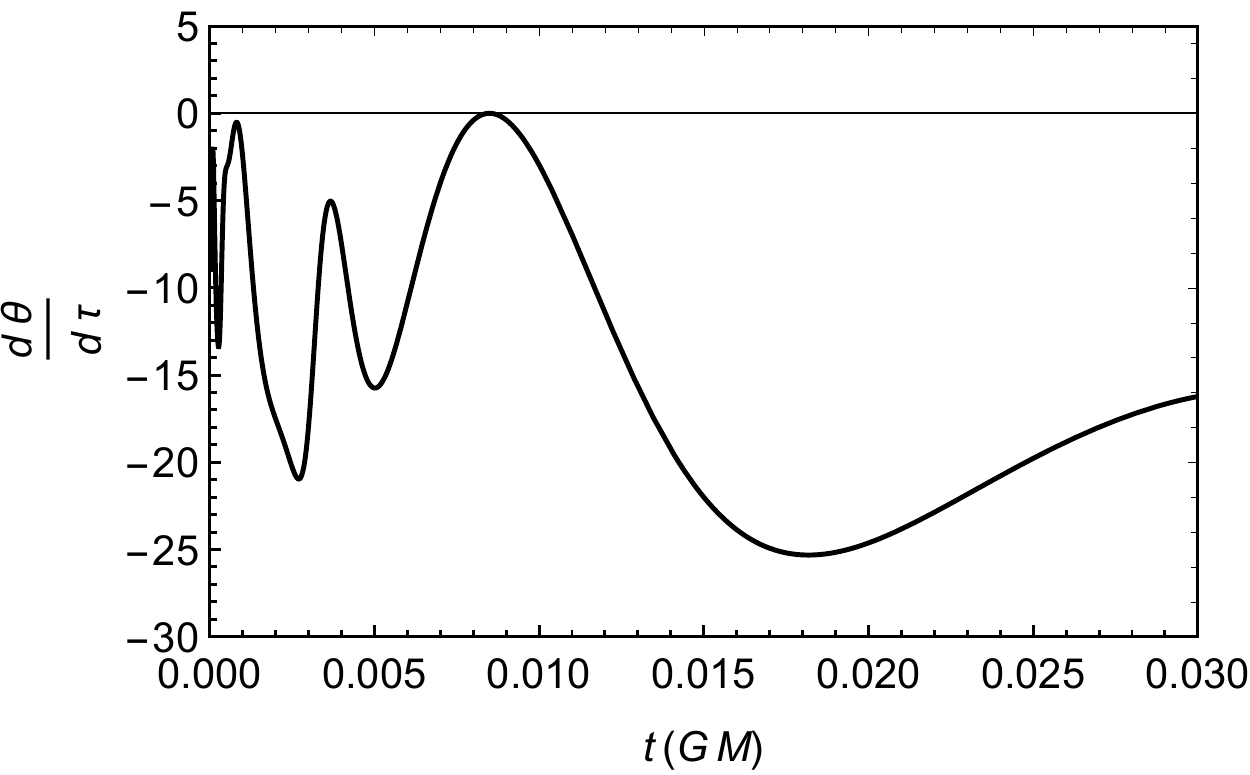}~~~~~~~~\includegraphics[scale=0.4]{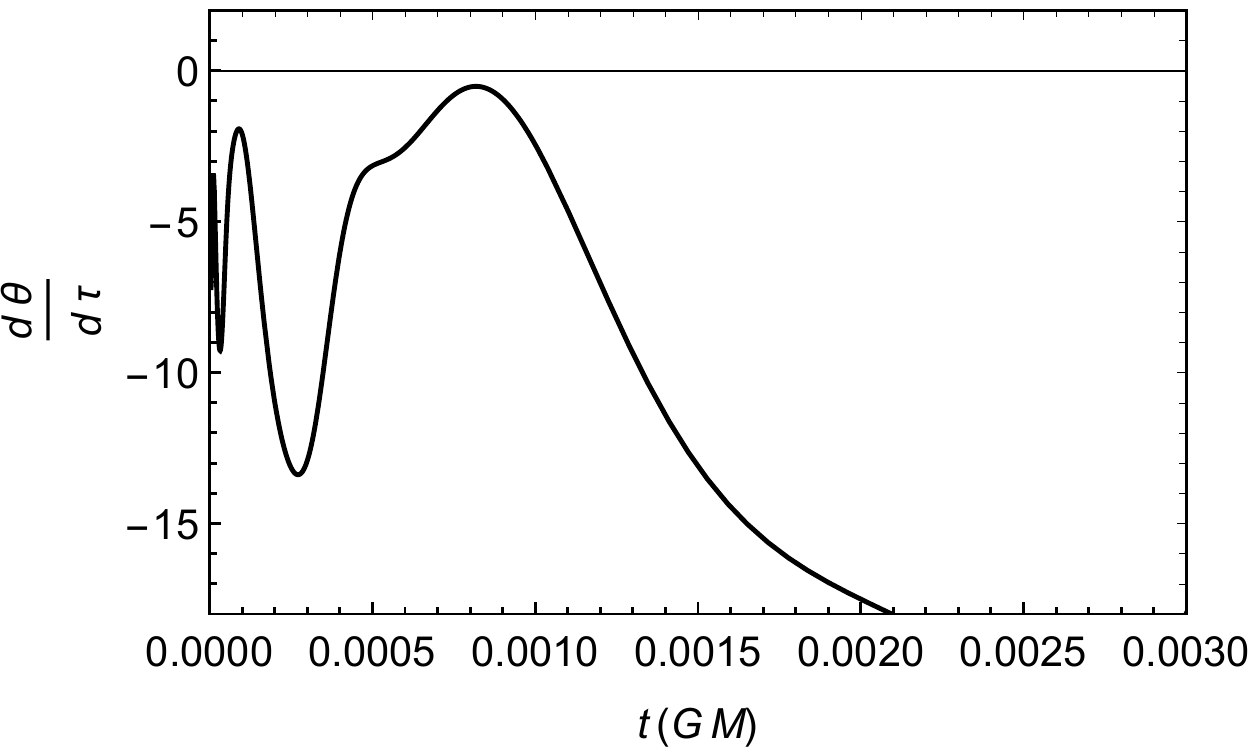} 
\par\end{centering}
\begin{centering}
\vspace{10pt}
 \includegraphics[scale=0.4]{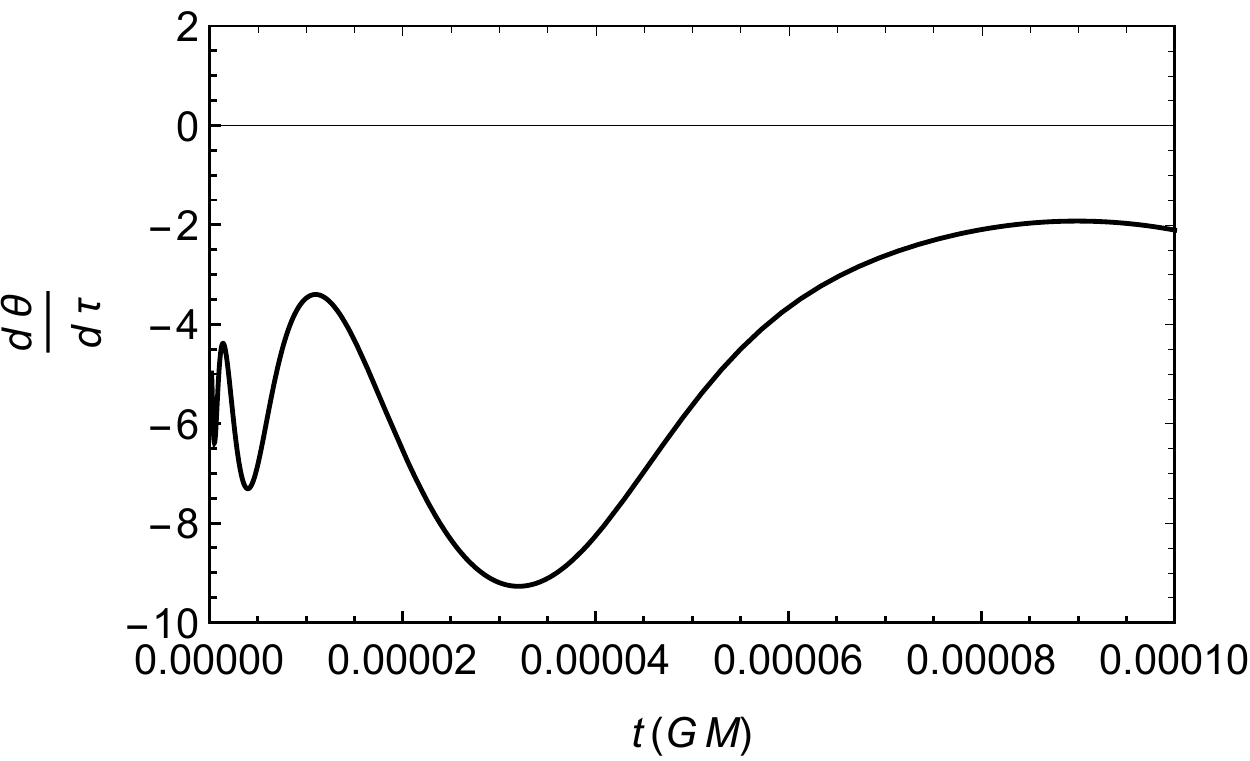}~~~~~~~~\includegraphics[scale=0.4]{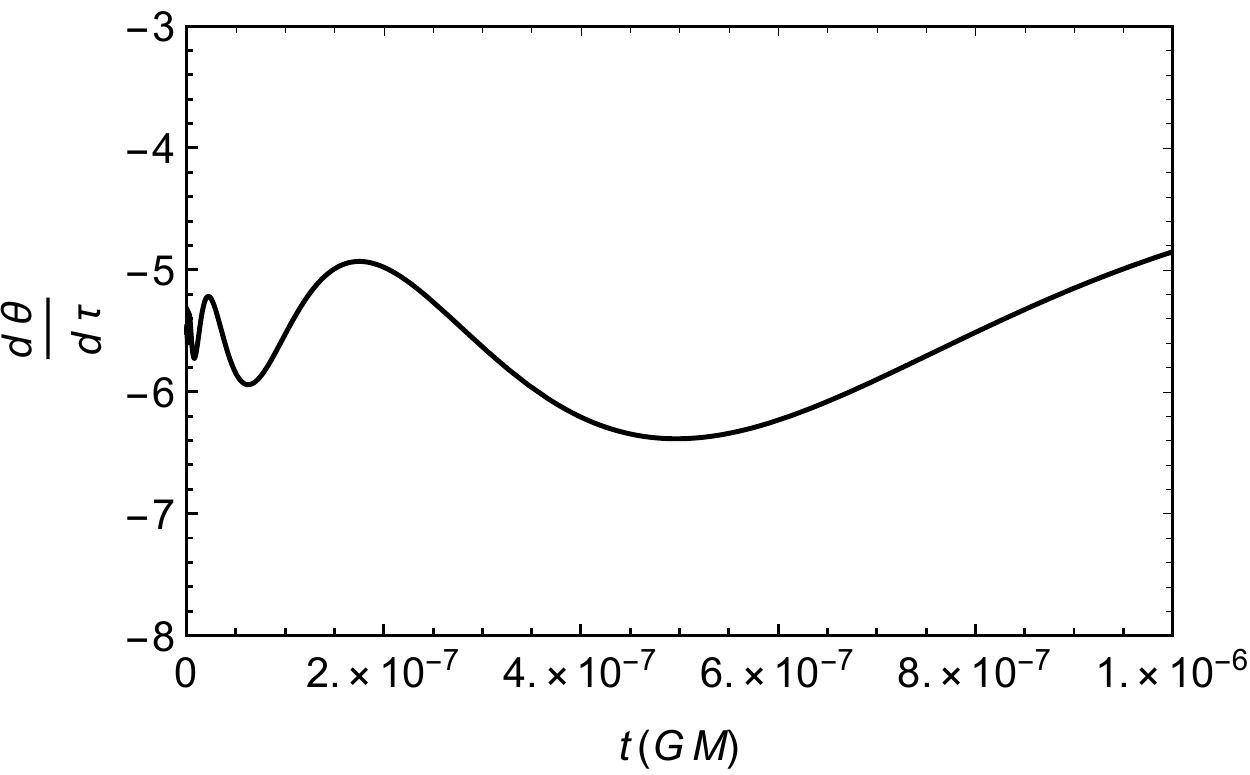} 
\par\end{centering}
\caption{Raychaudhuri equation in the $\bar{\mu}^{\prime}$ scheme. The top
left figure shows the behavior over the whole $0\protect\leq t\protect\leq2GM$
range. Other plots show various close-ups of that plot over smaller
ranges of $t$. The figure is plotted using $\gamma=0.5,\,M=1,\,G=1,\,L_{0}=1$,
and $\Delta=0.1$. \label{fig:RE-bar-prime}}
\end{figure}

The full nonperturbative Raychaudhuri equation and its close-ups in
this case are plotted in Fig. \ref{fig:RE-bar-prime}. It is seen
that in this scheme, the Raychaudhuri equation exhibits a more oscillatory
behavior and has various bumps particularly when we get closer to
where the singularity used to be. Very close to the classical singularity,
its form resembles those of $b$ and $p_{c}$, behaving like a damped
oscillation \citep{Blanchette:2020kkk}.

Two particular features are worth noting in this scheme. First, as
we also saw in previous schemes, quantum corrections kick in close
to the singularity and dominate the evolution such that the infinite
focusing is remedied, hence signaling the resolution of the singularity.
Second, this scheme exhibits a nonvanishing value for $\frac{d\theta}{d\tau}$
at, or very close to, the singularity. In Fig. \ref{fig:RE-bar-prime}
with the particular choice of numerical values of $\gamma,\,M,\,G,\,L_{0}$
and $\Delta$, the value of $\frac{d\theta}{d\tau}$ for $t\to0$
is approximately $-5.5$. Hence, although a nonvanishing focusing
is not achieved in this case at where the singularity used to be,
nevertheless, there exists a relatively small focusing and certainly
$\frac{d\theta}{d\tau}$ remains finite.

\section{Discussion and outlook\label{Sec:Disc}}

In this work, we probed the structure of the interior of the Schwarzschild
black hole, particularly the region close to the classical singularity,
using the effective Raychaudhuri equation. The effective terms in
this equation result from considering the effective modifications
to the Hamiltonian of the interior due to polymer quantization, which
is equivalent to loop quantization of this model. We found out that
while the classical rate of expansion $\frac{d\theta}{d\tau}$ diverges
for $r\to0$, the effective terms counter such a divergence close
to the singularity and make $\frac{d\theta}{d\tau}$ finite at $r\to0$.
We considered three main schemes of polymer quantization and the results
hold in all three. This is a strong indication that LQG points to
the resolution of the singularity in the effective regime.

It is also worth noting that very similar behavior has been derived recently for several cases of Generalized Uncertainty Principle (GUP) models \citep{Bosso:2020ztk,Blanchette:2020aca}. In particular, it seems that these cases bare a significant resemblance to $\mathring{\mu}$ and $\bar{\mu}$. This can be taken as a cross-model affirmation that quantum gravity in general does resolve the singularity of the Schwarzschild black hole. 

\bibliographystyle{ws-procs961x669}
\bibliography{mainbib}

\end{document}